\documentstyle[11pt,newpasp,twoside,epsf]{article}
\def\reference{\parskip 0pt\par\noindent\hangindent 0.5 truecm}
\def\kms{km ${\rm s}^{-1}$}

\def\edcomment#1{\iffalse\marginpar{\raggedright\sl#1\/}\else\relax\fi}
\reversemarginpar

\begin{document}
\title{An Extended View of the Fornax Cluster}
\author{M. Waugh, R. L. Webster, \& M. J. Drinkwater}
\affil{School of Physics, University of Melbourne, Victoria,
3010, Australia}

\begin{abstract}
The Multibeam survey at the Parkes Radio Telescope has provided a wealth 
of 21 cm HI data. We have mosaicked ten standard HIPASS cubes to 
produce a datacube approximately 25\deg$\times$25\deg\ in size, centred 
on NGC1399, the optical centre of the Fornax Cluster.  
Some properties of the initial $\sim$80 galaxies identified in HI 
are described. 

In the optical, Ferguson's Fornax Cluster Catalogue (FCC) (Ferguson 1989) 
identified 
340 likely member galaxies in the central 40 deg$^2$. In the radio 
we have detected a more uniform sheet of about 80 galaxies at the cluster 
velocity extending up to 15\deg\  from the cluster centre. 
At the cluster distance of about 15 Mpc, this corresponds to an elongated 
structure more than 7 Mpc in extent.

Galaxies were detected to a lower mass limit of $\sim$$1 \times 10{^8}$ 
M$_\odot$ and 
fewer than 25 of these were within the central 40 deg$^2$,
\hspace{2pt} suggesting considerable HI depletion of galaxies in the 
centre of the  cluster.  
Further, these results strongly indicate that HI surveys 
do {\it{not}} readily identify galaxy clusters.

\end{abstract}

\section{Introduction: The Fornax Cluster}

The Fornax Cluster is amongst the closest and most well studied clusters 
in the southern sky, providing a rich nearby field for the study 
of galaxy populations, dynamics and evolution in the cluster environment.

Covering an area of nearly 40 deg$^2$, Ferguson's optical 
catalogue (Ferguson 1989) identified 340 ``likely cluster members'', 
only 85 of which had redshifts.
Cluster membership was largely assigned on the basis of 
morphology and surface brightness (SB), suggesting a bias whereby high 
SB compact galaxies, paricularly unusual dwarf galaxies, may have been 
rejected as ``stars'' and any extended low surface brightness (LSB) galaxies 
incorrectly assigned a ``background'' status (Disney 1976).

Using the Tully-Fisher relation in Fornax, Bureau et al.\ (1996) 
determined the cluster distance to be \mbox{15.4 $\pm$ 2.3 Mpc}. 
Recessional velocities of Fornax cluster galaxies 
range from about 700 \kms\ to 2200 \kms\  (Barnes et al.\ 1997).  
Very  few galaxies 
have been detected in front of the cluster or in the void behind the 
cluster in the region \mbox{2200--4500 \kms}.  Galaxies beyond 
about \mbox{{\it{cz}} =  3000 \kms}  would not be classified as likely 
cluster members (Drinkwater \& Gregg 1998).

The fields covered by several surveys of the Fornax cluster are illustrated 
in Figure 1, including the unbiased spectroscopic Two Degree Field (2dF) 
Survey which is surveying  {\it{all}} objects in the magnitude limits 
16.5 $\leq$ $B_J$ $\leq$ 19.7 in a 
{\mbox{$\sim$12 deg$^2$}} region (Drinkwater et al.\ 2000a).  The positions of 
the 10 HIPASS cubes used to compile our mosaic are shown.  The deep neutral
hydrogen survey covering 10\deg$\times$10\deg\ in a ``basket-weave'' pattern 
has recently been completed  and new detections not found in HIPASS 
have already been identified (see Drinkwater et al.\ in these proceedings).

Also, a   spectrographic survey centred on Fornax 
using the FLAIR instrument on the UK Schmidt Telescope  
(not shown here) is identifying  bright compact objects over a  
6\deg$\times$6\deg\ 
field in this region (Drinkwater \& Gregg 1998).

\begin{figure}
\plotone{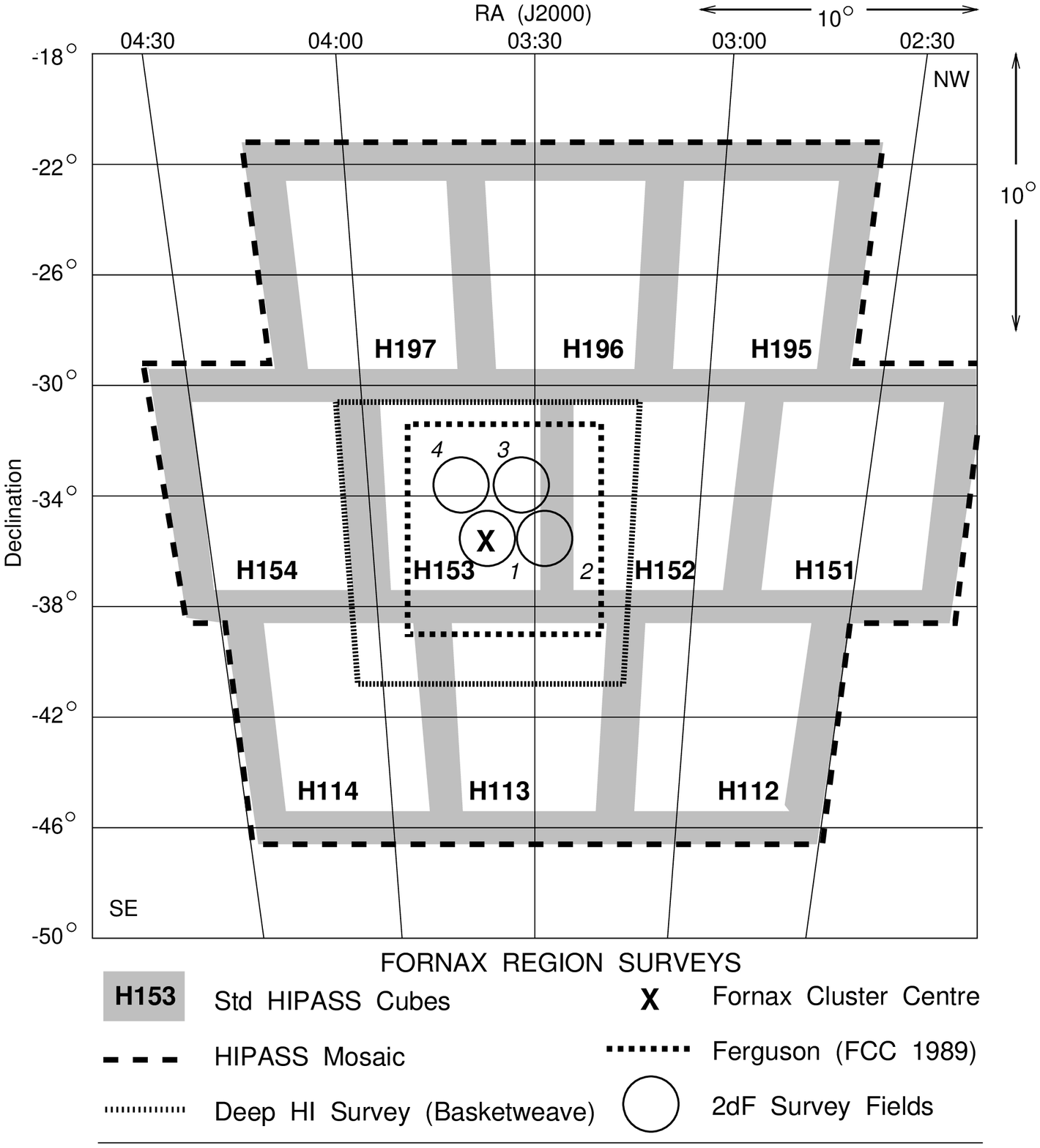}
\caption{Relative positions and approximate sizes of various surveys in the 
Fornax region.}
\end{figure}

\section{Our Data: HIPASS Mosaic}

Ten standard HIPASS cubes were used to create our 25\deg$\times$25\deg\ 
mosaic. 
HIPASS detects HI emission over the velocity range -1280$<cz<$12700 
\kms,  has an angular 
resolution of $\sim$15 arcmin, a velocity resolution of 
18 \kms\ and a 1$\sigma$ sensitivity of approximately 13 mJy beam$^{-1}$ 
(Barnes et al.\ 2000). 
At the distance of Fornax,  this corresponds  to a lower  
detectable HI mass limit of  {\mbox{$\sim10^8$ M$_\odot$}}.

After masking  the edges of the mosaic, an automated galaxy finder  
(Kilborn 2000) was applied to the data.  At a lower cutoff 
of 3$\sigma$ in peak flux (corresponding to about 33 mJy), 
the galaxy finder produced a first list of 
400 or so candidate detections by position and 
peak flux over a velocity range of 300--3700 \kms.\
 This  list was easily culled to remove known interference-generated 
``detections'' (derived from known telescope correlator interference) and 
to group duplicates together, leaving a list of 87 possible detections, each 
of which was confirmed by eye in the mosaic.

For each candidate detection, the {\sc{miriad}} software package was 
employed to establish an accurate position, spectral profile and other 
standard parameters.  The NASA databases NED and DSS were searched for 
likely optical counterparts to within a maximum 10 arcmin radius (although
positional matches could be expected to be within $\sim$6 arcmin).

We have confirmed approximately 80 HI detections, including several without an 
apparent 
optical counterpart, and several more with more than one object in the 
Parkes beam in the DSS images.   New velocity measurements have 
been determined for 8 previously identified galaxies.

Velocity histograms for 110 optical galaxies in the Fornax Cluster 
{\mbox(area  {$\sim$40 deg$^2$})} and for 78 HI detections in our mosaic 
(area $\sim$600 deg$^2$) are plotted in Figure 2. The optical identifications 
include both those listed in the Fornax Cluster Catalogue and those 
subsequently added by the FLAIR surveys (Drinkwater et al.\ 2000b).

\begin{figure}
\plotone{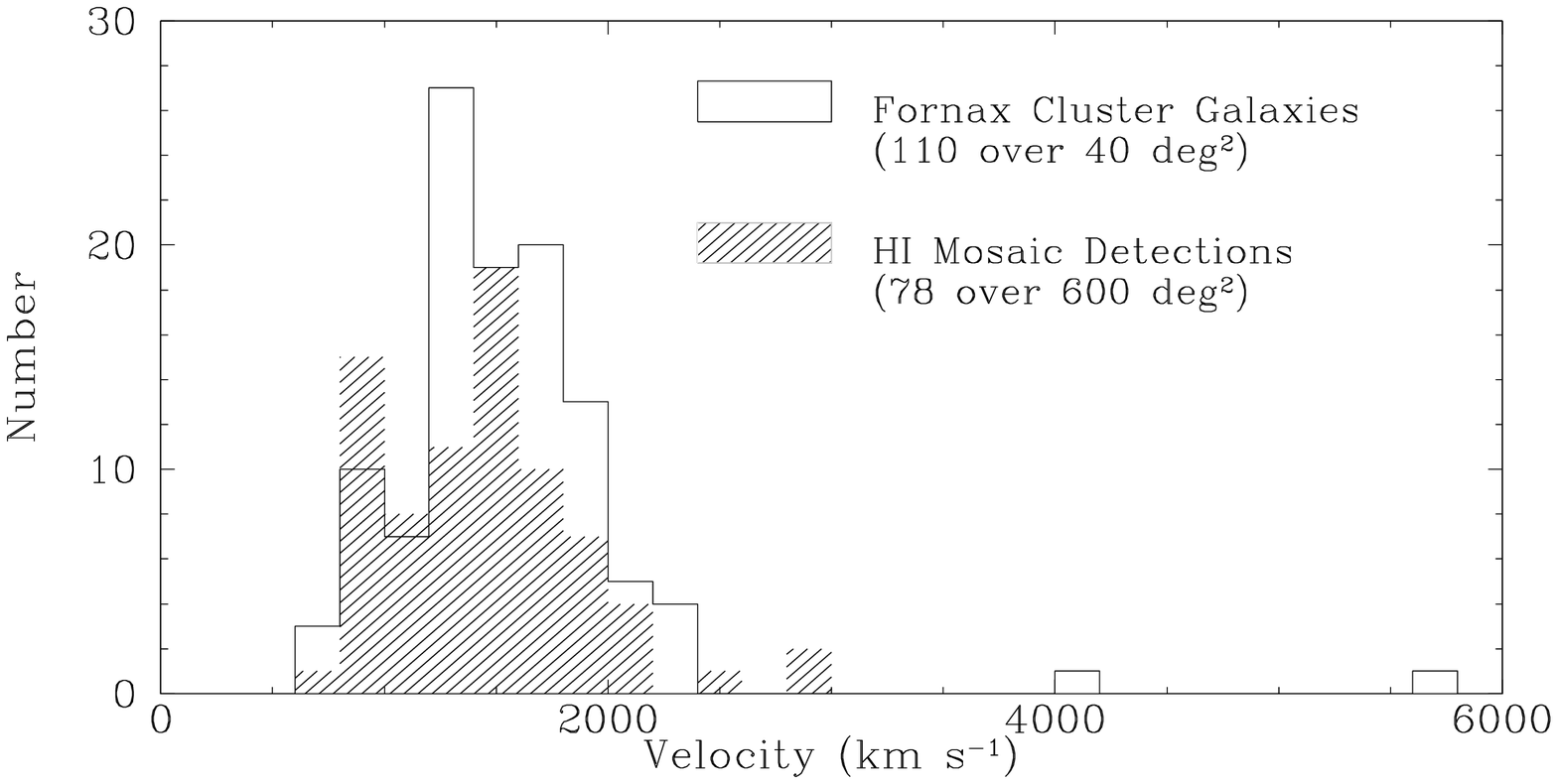} 
\caption{Velocity histograms for Fornax Cluster optical galaxies 
(solid lines) and for HI detections in our mosaic (shaded). Note that 
the 110 optical galaxies represented are within {\mbox{$\sim$40 deg$^2$,}} 
whilst the  
78 HI  detections span a much larger area. The void behind the cluster is 
evident in both.} 
\end{figure}

\section{Large-scale Structure around Fornax}

In Figure 3, the positions of the HI detections relative to the 
Fornax centre are shown.  
Large-scale structure (LSS) is also evident in Figure 4, where 
the galaxies appear to be in a long sheet-like arrangement in which 
the  velocity widths of the galaxies are comparable to the velocity 
dispersion of the whole group.  Our results indicate that there is a slight 
velocity gradient, from SE to NW, 
across the formation, as is suggested in Figures 3 and 4.

\begin{figure}
\plotone{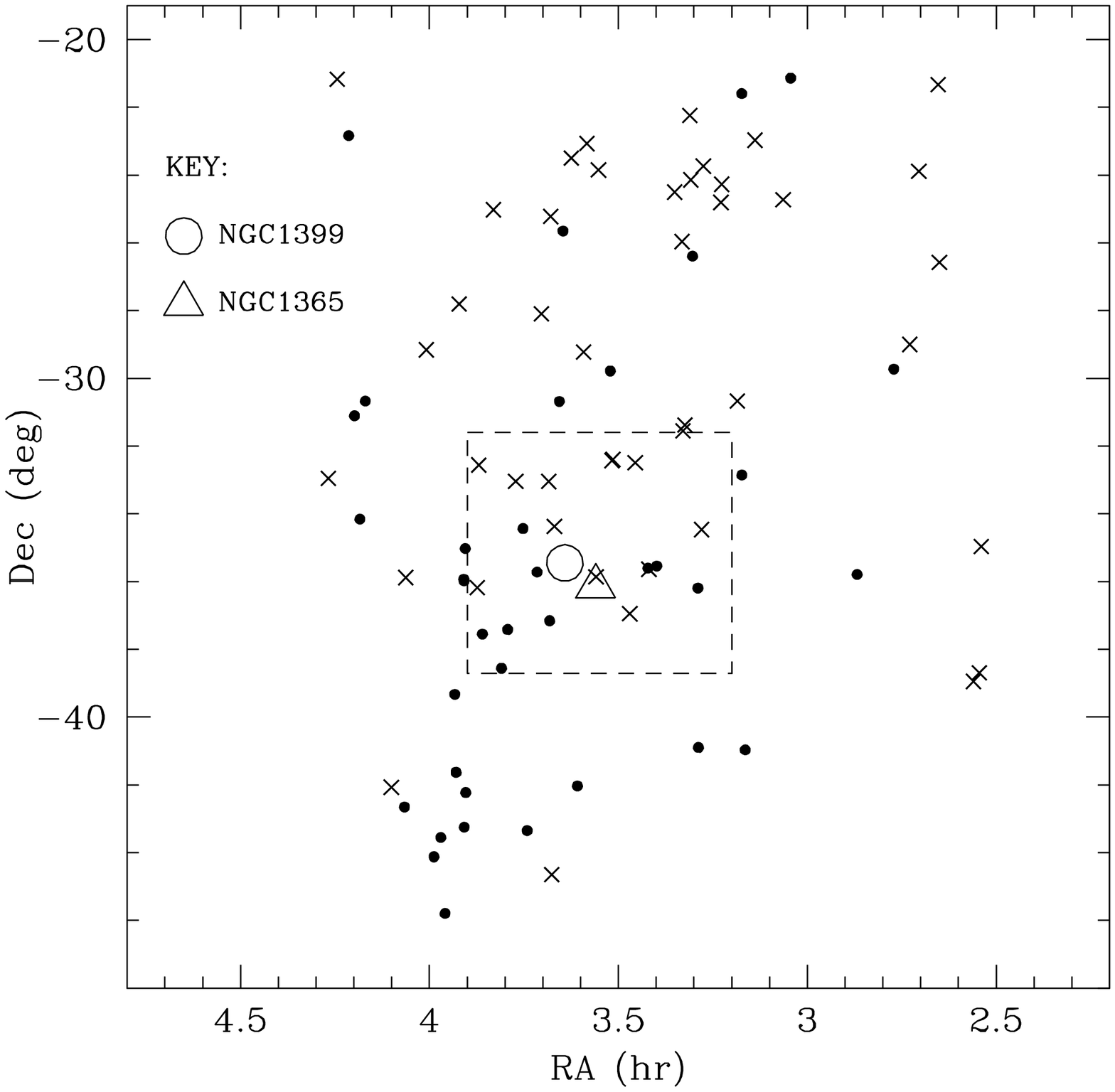} 
\caption{Positions of Fornax Mosaic HIPASS detections. The approximate 
boundary of the Fornax Cluster Catalogue is 
indicated (dashed line). The  position of the giant elliptical NGC1399, at a 
recessional velocity of 1425 \kms , is 
generally adopted as the cluster centre.  The positions of NGC1399 
(a non-detection) and NGC1365 (a large spiral) are shown. 
Detections at $<$1400 \kms \ are shown
as dots, those at $>$1400 \kms \ are shown as crosses.  A velocity gradient 
across the structure from SE to NW is suggested.} 
\end{figure}

\begin{figure}
\plotone{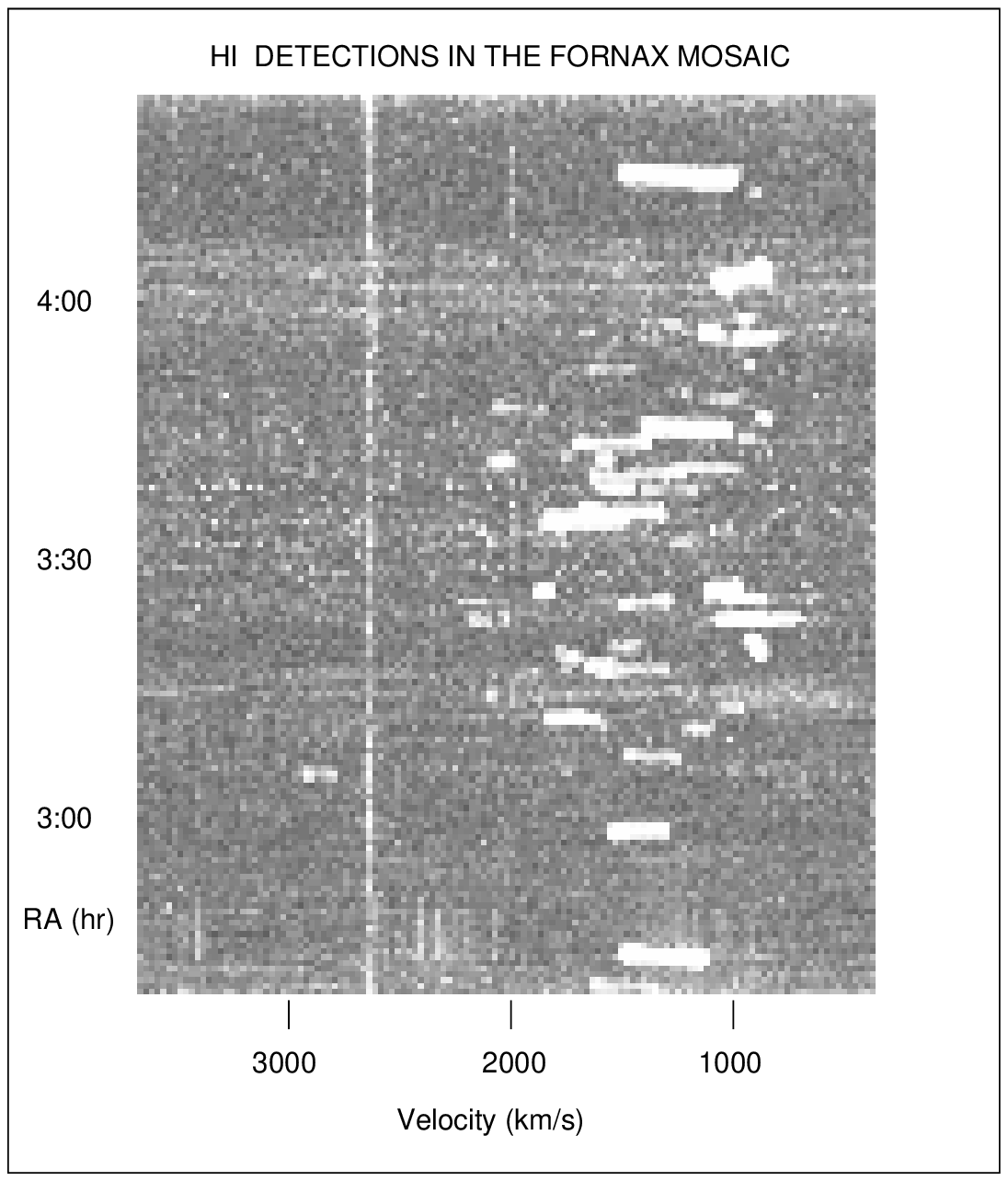}
\caption{HI flux map from the Fornax mosaic showing velocities, velocity 
widths and RA.  Detections appear as bright bands. The interference 
band around 2630 \kms\ ($\sim$1408 MHz) (see Section 2 above), is seen 
as the  bright line present through all RA positions.} 
\end{figure}

HIPASS is a survey biased towards the detection of HI-rich objects. As is 
evident in Figure 3, the Fornax cluster of galaxies at the centre of the 
mosaic field is not clearly distinguished in this survey. Only 5 
(less than 7\%) of the HI detections are within the 
central 2.4\deg\ radius core of the cluster and fewer than about 30\% 
of detections are within the boundary of the Fornax Cluster Catalogue region.
This would suggest considerable HI depletion in the cluster galaxies, possibly 
due to tidal stripping, ram pressure stripping, star formation  and/or 
other mechanisms.

Our results clearly indicate that HIPASS is {\it{not}} 
a survey which will preferentially detect galaxy clusters.

\section{HI Mass Distribution}

The HI mass ({\mbox{M$_{\rm HI}$}})  of each detection 
was estimated using  the standard formula (Verschuur \&  Kellermann 1988). 
For {\mbox{M$_{\rm HI}$}} calculations, all detections were assumed to be at 
the cluster distance of 15.4 Mpc. 
Where known, the blue magnitude values ($m_B$) given in NED were used for 
mass-to-light ratio ({\mbox{M$_{\rm HI}$}/L}) 
estimations.  Again, all detections were assumed 
to be at the same distance.  Note that the NED values have errors of 
$\leq$0.5 $m$ 
in the total blue magnitude (Marquarding 2000).

At least 60\% of our initial detections have M$_{\rm HI}$ less 
than $10^9$ M$_\odot$ and as many as 50\% are dwarfs with an HI mass of 
less than 5$\times$10$^8$ M$_\odot$.

In Figure 5 it can be seen that it is predominantly the low mass galaxies  
which have  M$_{\rm HI}$/L ratios greater than 1--2 M$_\odot$/L$_\odot$. 
The higher mass 
galaxies are presumably depleted of HI by a higher level of past or 
current star formation.

\begin{figure}
\plotone{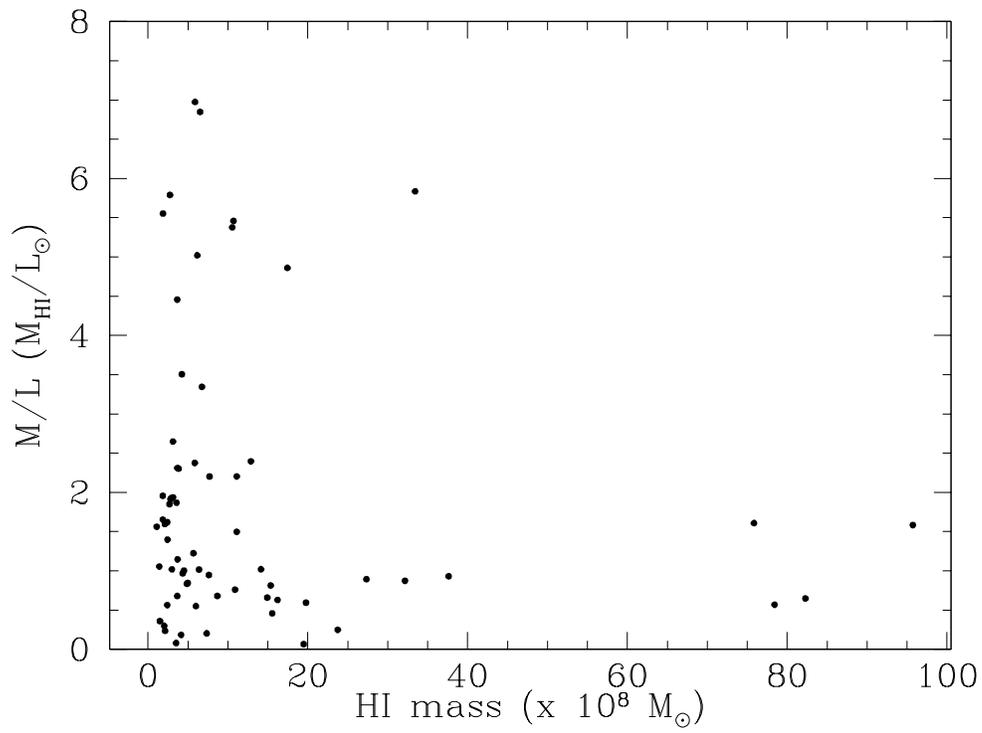}
\caption{In this plot of M$_{\rm HI}$ vs. {\mbox{M$_{\rm HI}$}/L}, more 
than 60\% of detections have an HI mass less than $10^9$ M$_\odot$. Generally,
only the small or dwarf galaxies appear to have high {\mbox{M$_{\rm HI}$}/L}
ratios.}
\end{figure}

\section{Summary}
A mosaic of 10 standard HIPASS cubes, spanning about 25\deg$\times$25\deg\  
and centred on the Fornax Cluster, has been created 
and efficiently searched using an automated galaxy finder.

Fewer than about 30\% of the detections are within the central 
40 deg$^2$ of the 
cluster, suggesting considerable HI depletion of galaxies in the cluster.  
Further, our results show that HI surveys  
do not readily detect  galaxy clusters.

Down to a lower M$_{\rm HI}$ limit of $10^8$  M$_\odot$, many low HI mass 
galaxies were detected. Initial results indicate 
that generally only low {\mbox{M$_{\rm HI}$}} galaxies have high HI 
mass-to-light ratios ({\mbox{M$_{\rm HI}$}/L}).
New velocities of at least 8 objects have been measured. 
Interestingly, several HI detections do not have optical counterparts.

Our initial $\sim$80 confirmed HI detections have revealed that the 
Fornax Cluster is 
imbedded in a large-scale structure  more than 7 Mpc in extent and 
evidence of a velocity gradient across the LSS is seen.

\acknowledgments

Many thanks to Virginia Kilborn for her generous assistance and use of 
her ``galaxy finder'' program.

\end{document}